# Doppler-Free Two-Photon Cavity Ring-Down Spectroscopy of a Nitrous Oxide (N₂O) Vibrational Overtone Transition

Gang Zhao,[1,†] D. Michelle Bailey,[1] Adam J. Fleisher,[1,*] Joseph T. Hodges,[1] and Kevin K. Lehmann[2]

[1]*National Institute of Standards and Technology, 100 Bureau Drive, Gaithersburg, Maryland 20899*

[2]*Departments of Chemistry and Physics, University of Virginia, Charlottesville, Virginia 22904*

[†]*Permanent address: Institute of Laser Spectroscopy, State Key Laboratory of Quantum Optics and Quantum Devices, Shanxi University, Taiyuan City 030006, Shanxi Provence, P. R. China*

[*]adam.fleisher@nist.gov

Dated: 27 March 2020

*Abstract*—We report Doppler-free two-photon absorption of N$_2$O at $\lambda$ = 4.53 μm, measured by cavity ring-down spectroscopy. High power was achieved by optical self-locking of a quantum cascade laser to a linear resonator of finesse $\mathcal{F}$ = 22730, and accurate laser detuning over a 400 MHz range was measured relative to an optical frequency comb. At a sample pressure of $p$ = 0.13 kPa, we report a large two-photon cross-section of $\sigma_{13}^{(2)}$ = 8.0 × 10$^{-41}$ cm$^4$ s molecule$^{-1}$ for the $Q$(18) rovibrational transition at a resonant frequency of $\nu_0$ = 66179400.8 MHz.

*Introduction*—Two-photon spectroscopy of the 1*S*-2*S* transition of hydrogen (H) has inspired 45 years of precision laser measurements [1, 2]. First reported by Hänsch et al. in 1975 [3], the fractional uncertainty in this transition frequency is currently 4.2 × 10$^{-15}$ [4], a record enabled by remarkable advances in tools for frequency metrology and control like the optical frequency comb [1, 5]. In general, two-photon absorption [6, 7] by counter-propagating beams of identical frequency eliminates the first-order Doppler effect [8, 9], resulting in ultra-narrow lines with homogeneous broadening ultimately limited by the natural lifetime of the upper state—ideal for precision spectroscopy. Consequently, two-photon spectroscopy remains at the forefront of several challenges in modern physics [10], including tests of charge-parity-time (CPT) symmetry using antihydrogen [11] and attempts to solve the proton-size puzzle [12].

With additional degrees of freedom, molecules possess two-photon absorption lines at frequencies outside of the ultraviolet and visible spectrum. However, only molecules with fortuitous transition frequencies near the emission lines of high-powered gas lasers have been observed [13-15]. Because infrared cross-sections are generally smaller than those associated with electronic transitions, new techniques with substantially greater sensitivity are required. Nevertheless, several intriguing tests of fundamental physics enabled by molecules await, like the search for parity violations in chiral species [16] and tests of fundamental constants by hydrogen molecular ions [17]. The theories associated with these tests have motived recent advances in tunable and coherent infrared radiation with state-of-the-art frequency accuracy [18, 19] using emerging laser sources [20].

Here we report Doppler-free two-photon absorption of the $Q(18)$ rovibrational transition within the $\nu_3$ overtone band of $N_2O$, measured by cavity ring-down spectroscopy (CRDS) at a wavelength of $\lambda = 4.53$ μm. As proposed by Lehmann in Ref. [21], we demonstrate two-photon CRDS to be a sensitive probe for selected light-matter interactions within an optical cavity. Resonance enhanced two-photon CRDS has a distinct sensitivity advantage over saturated absorption techniques [22-25], where the former Doppler-free process benefits from a near degeneracy of energy levels and the fact that all velocity classes contribute to the two-photon cross-section. Here we directly measured the resonance enhanced two-photon absorption rate for a chosen $N_2O$ transition, as well as its transition frequency, pressure shift, and collisional air-broadening coefficient. Furthermore, we project an ultimate $N_2O$ detection limit for our phase-locked quantum cascade laser (QCL) spectrometer which promises a 125-fold improvement over the best-available laser gas analyzers using cavity-enhanced linear absorption techniques.

These measurements required the creation of a novel cavity-locked QCL spectrometer with a tunable frequency axis that was referenced to a stabilized optical frequency comb. The spectrometer, with ultra-narrow relative linewidth and high circulating power, comprised a semiconductor laser controlled by an optical-electronic phase-locking scheme.

*Results*—The experimental setup shown in Fig. 1(a) was developed and implemented at the National Institute of Standards and Technology (NIST) in Gaithersburg, Maryland, USA. The continuous-wave distributed feedback QCL (HHL-14-75, AdTech Optics) with anti-reflection coated output facet emitted at $\lambda = 4.53$ μm with a collimated output power of ~100 mW. The QCL drive current was modulated by a sinusoidal signal at 1.8 MHz to create sidebands for electronic stabilization of the round-trip feedback phase. A pair of lenses, i.e. lens$_1$ and lens$_2$, were used to perform spatial mode-matching to a Fabry-Pérot type linear optical resonator (or cavity). The cavity comprised two plano-concave mirrors with 1 m radius of curvature separated by a stainless-steel vacuum enclosure with invar support rods, resulting in an optical single-pass cavity length of nominally $L = 75$ cm, free spectral range of 200 MHz, and beam waist radius at the focus of 0.835 mm.

QCL locking to the cavity was realized by optical feedback from the cavity leak-out field in a set-up similar to Ref. [26] and without the need for intentional mode-mismatching [27-29]. The feedback ratio could be adjusted without loss of incident power to the cavity by rotating a quarter-wave plate ($\lambda/4$) to change the deflection ratio of the reflected light at the polarizing beam splitter (PBS). Stable feedback was established when the fast axis of $\lambda/4$ was rotated ~45° relative to the transmission axis of the PBS, resulting in circularly polarized light entering the cavity and an estimated feedback fractional intensity of ~$10^{-4}$. The feedback phase condition was initially satisfied by adjusting the length between the laser and the cavity to approximately equal the cavity length $L$. Then, electronic phase stabilization was added by controlling the laser-cavity path length using a piezo-mounted mirror (PZT$_1$) in the light propagation path driven by a proportional-integral-derivative servo with a cascaded double-integral circuit (D2-125, Vescent Photonics). The error signal was generated in a manner similar to Ref. [30], i.e. by demodulating the reflected light deflected by PBS and received by photodetector PD$_2$.

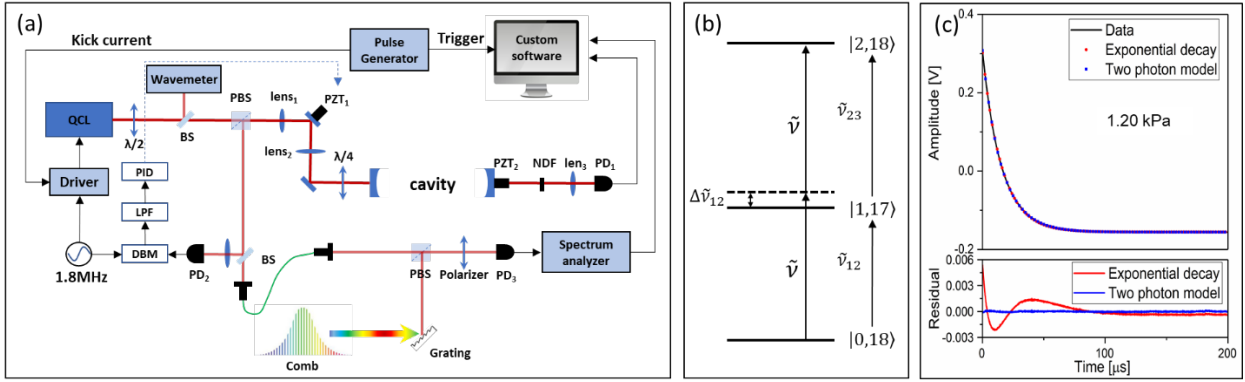

**Fig. 1(a).** Experimental setup. QCL, quantum cascade laser; $\lambda/2$, half-wave plate; $\lambda/4$, quarter-wave plate; BS, beam splitter; PZT, piezo-electric transducer; NDF, neutral density filter; PD, photodetector; DBM, double-balanced mixer; LPF, low-pass filter; PID, proportional-integral-derivative. **(b)** Resonance enhanced Q(18) two-photon transition, $\nu_3$ vibrational ladder of $N_2O$. Quantum states, $|\nu_3, J\rangle$ **(c).** Upper panel. Cavity ring-down signal, 24.8 μmol mol$^{-1}$ $N_2O$-in-air, pressure $p$ = 1.20 kPa (black line). Exponential (red dots) and two-photon (blue dots) fitted models. Lower panel. Fitted residuals.

The mirror power reflectivity ($\mathcal{R}$), transmittivity ($\mathcal{T}$), and combined loss coefficients ($\mathcal{L}$, absorption and scattering), as well as the fraction of the incident power coupled into the lowest order transverse cavity mode ($\varepsilon$) were estimated by measuring the empty-cavity ring-down time, the optical power reflected from the cavity both on- and off-resonance, the incident power, and the transmitted on-resonance power. From those measurements, we inferred the following mirror and spectrometer parameters [31, 32]: $\mathcal{R}$ = 0.9998618, $\mathcal{T}$ = 6.99 × 10$^{-5}$, $\mathcal{L}$ = 6.83 × 10$^{-5}$, and $\varepsilon$ = 0.54. The calculated cavity finesse was 22730 with a standard uncertainty of 160. On resonance and at an incident power of 26 mW, the empty-cavity transmitted power was measured to be 4.1 mW, which implies a one-way intracavity power of 60 W.

Cavity decay signals were measured in transmission by a liquid-nitrogen-cooled InSb photoconductive detector (PD$_1$, responsivity of 3.5 A/W) following a neutral density filter (NDF) with optical density OD = 1.2. Detector response was converted to voltage by a transimpedance amplifier and then digitized by an oscilloscope card (PCI5922, National instruments) with 18-bit resolution at a sampling rate of $10^7$ s$^{-1}$. The cavity-locked laser frequency could be continuously tuned over a range of 400 MHz by changing the voltage to a piezo-electric transducer (PZT$_2$) attached to one of the cavity mirrors. Low and high precision values for the laser wavelength were determined, respectively, by a wavelength meter (721, Bristol) and by beating with a stabilized frequency comb (Menlo Systems). The frequency comb was an offset-free difference frequency generation system ($f_0$ = 0) with repetition rate $f_{rep}$ = 250 MHz when locked to a Rb clock signal (SIM940, Stanford Research Systems, short-term stability of <2 × 10$^{-11}$ at 1 s). The beat note frequency was recorded by a spectrum analyzer and processed by a custom software program to calculate the laser frequency at each point of the two-photon spectra relative to the comb.

Decay events were triggered by rapidly changing the QCL drive current, thus breaking the optical-electronic phase-locked loop. The two-photon transition of $^{14}N_2^{16}O$ at a calculated wave number of $\tilde{\nu}_0$ = 2207.507 cm$^{-1}$ was chosen because of strong resonance enhancement between the $P(18)$ transition of the $\nu_3$ fundamental and the $R(17)$ transition of the first $\nu_3$ vibrational hot band (Fig. 1(b)). The intermediate state is detuned by only 0.113 cm$^{-1}$, i.e. 3.39 GHz, from the predicted resonant frequency of the two-photon transition, resulting in a calculated two-photon cross-section at 296 K and 0.13 kPa of $\sigma_{13}^{(2)} = 5.9 \times 10^{-41}$ cm$^4$ s molecule$^{-1}$. Moreover, the two-photon absorption is detuned by ~900 MHz from the nearest one-photon transition of N$_2$O [33].

The gas sample comprised 24.8069 µmol/mol ± 0.0076 µmol/mol N$_2$O in air (expanded uncertainty, or coverage factor $k = 2$, cylinder # FB03344), and was gravimetrically prepared at NIST using synthetic air of largely O$_2$ and N$_2$ [34]. Sample temperature ($T_g$) and pressure ($p$) were measured by two platinum resistance thermometers in good thermal contact with the outside of the sample cell and a 1.3 kPa full-scale pressure gauge, respectively, each of which was calibrated against secondary NIST standards.

The black line in Fig. 1(c) shows the measured two-photon ring-down signal at 1.20 kPa of N$_2$O-in-air, triggered near the center of the $Q(18)$ $\nu_3$ transition. Red and blue dots show the fitted exponential decay and two-photon cavity ring-down models, respectively. The two-photon ring-down model [21, 23] is reproduced below in Eq. (1), and the fit in Fig. 1(c) involved floating the following parameters: $\gamma_1$, $\gamma_2$, the voltage at the detector at zero time $V_{\text{det}}(0)$, and a detector offset $V_0$.

$$V_{\text{det}}(t) = \frac{\gamma_1 V_{\text{det}}(0) \exp(-\gamma_1 t)}{\gamma_1 + \gamma_2 \left(\frac{V_{\text{det}}(0)}{G_d \mathcal{T}}\right)(1-\exp(-\gamma_1 t))} + V_0 \qquad (1)$$

The quantity $V_{\text{det}}(0)/(G_d \mathcal{T})$ is equal to the intracavity power at zero time $P_{\text{ic}}(0)$ in units of W, where $G_d = 215$ V W$^{-1}$ is the measured gain of the detection system including the transmittance associated with the neutral density filter (NDF) shown in Fig. 1(a). Also in Eq. (1), $V_{\text{det}}(t)$ is the voltage measured by the detection system as function of time ($t$) in units of V, $\gamma_1$ is the one-photon absorption rate (which includes both mirror losses and linear molecular losses) in units of s$^{-1}$, and $\gamma_2$ is the two-photon absorption rate in units of s$^{-1}$ W$^{-1}$. Shown in the lower panel of Fig. 1(c) are the fitted residuals corresponding to each model. Systematic structure in the residuals of the exponential decay model, with maximum deviation of up to 2% of $V_{\text{det}}(0)$, illustrates the presence of strong non-linear absorption. In contrast, fitting with two-photon ring-down model yielded low root-mean-square (RMS) noise. Defining the ring-down signal-to-noise ratio (SNR) as the signal at zero time divided by the RMS noise, we observed SNR = $6.5 \times 10^3$ for one decay event.

Two-photon spectrum of N$_2$O over the frequency interval of ±200 MHz about the center of the transition was acquired by adjusting the cavity PZT$_2$ voltage to tune the phase-locked laser frequency. To eliminate correlation between $\gamma_1$ and $\gamma_2$ observed while fitting individual two-photon decay signals (correlation coefficient, $\rho_{12} \approx -0.95$), we constrained $\gamma_1$ to be a linear function of frequency detuning. The linear constraint captured frequency-dependent losses from the far wing of the $P(18)$ fundamental transition as well as spatially dependent changes in mirror coating losses. We also fixed the initial voltage $V_{\text{det}}(0)$ to equal the first data point in each decay and fixed the offset term $V_0$ to equal the average of the final ten data points in the decay. Therefore, $\gamma_2$ was the only model parameter floated in Eq. (1) during spectral analysis.

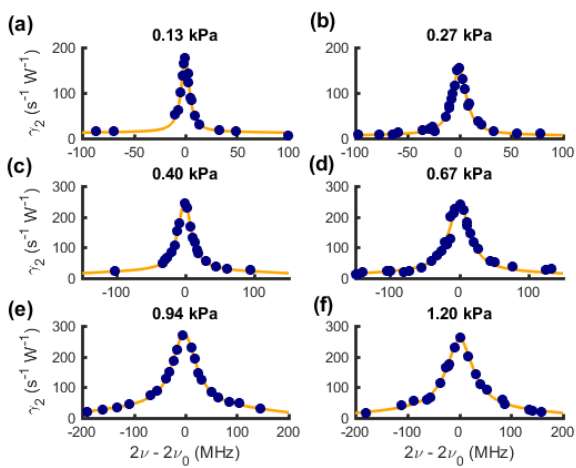

**Fig. 2.** Measured $\gamma_2$ as a function of two-photon detuning $2\nu - 2\nu_0$ for the N$_2$O $Q(18)$ $\nu_3$ overtone transition. Panels labeled by $p$. Resonant two-photon transition frequency, $\nu_0 = 66179400.8$ MHz $\pm$ 0.3 MHz.

Two-photon spectra, recorded at six different gas sample pressures from 0.13 kPa to 1.20 kPa, are shown in Fig. 2 (dark blue dots), along with their corresponding fitted models (orange lines, Lorentzian line shape function, Doppler-broadened pedestal [2]). The resonant two-photon transition frequency was calculated at zero pressure from a linear fit of the data in Fig. 2: $\nu_0 = 66179400.8$ MHz $\pm$ 0.3 MHz. Our measurement agrees with the HITRAN2016 value but is reported here with a tenfold lower statistical uncertainty.

Pressure-dependent values for the resonant two-photon transition frequency ($\nu_{0,p}$) are plotted versus $p$ in Fig. 3(a). Throughout Fig. 3, expanded ($k = 2$) statistical uncertainties from the fitting process are shown. The transition frequencies are reported relative to the value at zero pressure, i.e. $\nu_{0,p} - \nu_0$, where $\nu_0$ and the pressure shift coefficient $\delta_{air}$ resulted from a weighted linear fit, i.e., the black dashed line. The measured pressure shift for the resonant frequency of the two-photon transition is $\delta_{air} = -0.6$ MHz kPa$^{-1}$ $\pm$ 0.8 MHz kPa$^{-1}$, comparable to the average pressure shift reported in HITRAN2016 [33] for the two near-degenerate one-photon transitions, $\delta_{air} = -0.63$ MHz kPa$^{-1}$.

The fitted two-photon homogeneous broadening ($\Gamma_{air}$) as a function of $p$ is plotted in Fig. 3(b). The orange line, with fitted air-broadening coefficient $b_p = 25.7$ MHz kPa$^{-1}$ $\pm$ 1.2 MHz kPa$^{-1}$, includes an estimate for two-photon saturation [21]. For comparison, the air-broadening coefficients (half-width at half-maximum) for both the near-resonant one-photon transitions are $b_p^{(1)} = 22.5$ MHz kPa$^{-1}$ [33]. While fitting the orange line, $\Gamma_{air}$ at zero pressure was fixed to be twice the sum of our absolute laser linewidth (~270 kHz at 1 s) and the calculated transit-time broadening (30 kHz). Collisional air-broadening without saturation is predicted by the black dashed line.

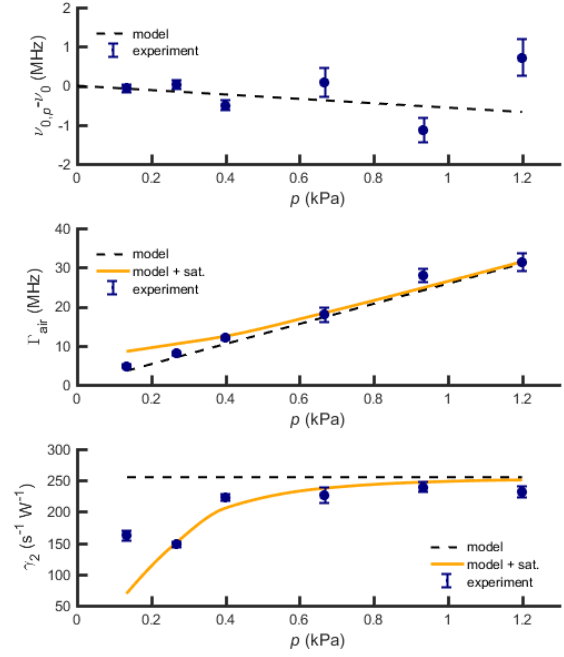

**Fig. 3.(a).** Center frequencies $\nu_{0,p}$ from fits of the two-photon cavity ring-down spectra vs. $p$. Black dashed line, weighted linear fit. **(b).** Two-photon half-width at half-maximum ($\Gamma_{\text{air}}$) vs. $p$. Orange line, fitted air-broadening coefficient including power-broadening. Black dashed line, model without power-broadening **(c).** Two-photon absorption rate ($\gamma_2$) vs. $p$. Orange line, fitted $\gamma_2$ including saturation. Black dashed line, fitted value of $\gamma_2$ without saturation.

Our transition frequency and collisional air-broadening analyses ignored unresolved nuclear quadrupole hyperfine structure attributable to each nitrogen atom, as well as any AC Stark shift. Further, we also ignored potential quantum-interference effects and one-photon saturation, including from the far-wing of the one-photon $P(18)$ $\nu_3$ fundamental transition which participates in the two-photon resonance enhancement.

Values of $\gamma_2$ fitted at $\nu_0$ (dark blue dots) are plotted in Fig. 3(c) along with the theoretical values calculated from Eq. (2) (black dashed line).

$$\gamma_2 = \frac{A_{12} A_{23} a_p(J_1, J_2, J_3) \chi_a f_1}{256 \pi^4 h c^3 L k_B T_g \Delta \tilde{\nu}_{12}^2 \tilde{\nu}^4 b_p} \tag{2}$$

Details regarding the derivation of Eq. (2) are available in Ref. [21]. Here, we summarize the parameters relevant to our experimental observations, noting that Eq. (2) was derived in the limit of a confocal cavity and is valid only when the near degeneracy ($\Delta \tilde{\nu}_{12} = \tilde{\nu}_{12} - \tilde{\nu}$) is much greater than both the one-photon Doppler and collisional broadening. In Eq. (2), $A_{12}$ and $A_{23}$ are the

-6-

Einstein coefficients for spontaneous emission, $a_p$ is a polarization-dependent factor calculated from Table II of Ref. [21], $\chi_a$ is the known mole fraction of absorbers, $f_1$ is the Boltzmann factor for selected isotopologue or isotopomer in its lower state, $h$ is the Planck constant, $c$ is the speed of light, $L$ is the known cavity length, $k_B$ is the Boltzmann constant, $T_g$ is the measured gas temperature, $\tilde{\nu}$ is the resonant wave number of the two-photon transition, and $b_p$ is the measured two-photon collisional air-broadening coefficient (Fig. 3(b)).

At higher pressures in Fig. 3(c), the experimental values of $\gamma_2$ approach a constant value of 255 s$^{-1}$ W$^{-1}$ ± 7 s$^{-1}$ W$^{-1}$. The orange line shows the fitted $\gamma_2$ including saturation of the two-photon transition, i.e. multiplying Eq. (2) by the factor of $1/(1 + G_{TP}^2)$, where $G_{TP}$ is the degree of two-photon saturation $G_{TP} = P_{ic}/P_{sat}$. In Ref. [21], $P_{sat}$ for the two-photon transition is predicted to be proportional to $p$ in the collisional broadening regime. However, that prediction may over-estimate the saturation power due to its neglect of expected very slow vibrational-to-translational and vibrational-to-rotational relaxation (e.g., [35]). The fitted two-photon absorption parameters for the N$_2$O $Q(18)$ $\nu_3$ overtone transition are summarized in Table I.

**Table I.** Summary of fitted two-photon absorption parameters along with database calculations using HITRAN2016 data [33] and equations in Ref. [21]. Two-photon absorption cross-sections ($\sigma_{13}^{(2)}$) were calculated per molecule of N$_2$O and correspond to $p = 0.13$ kPa.

| Parameter | Units | Measured value | Statistical uncertainty ($k = 1$) | Database calculation |
|---|---|---|---|---|
| $\nu_0$ | MHz | 66179400.8 | 0.3 | 66179401 |
| $\delta_{air}$ | MHz kPa$^{-1}$ | −0.6 | 0.8 | −0.63 |
| $b_p$ | MHz kPa$^{-1}$ | 25.7 | 1.2 | 22.5 |
| $\sigma_{13}^{(2)}$ | cm$^4$ s molecule$^{-1}$ | $8.0 \times 10^{-41}$ | $0.2 \times 10^{-41}$ | $5.9 \times 10^{-41}$ |

At $p = 1.20$ kPa, where $\gamma_2$ approaches the asymptotic value at high $p$ shown in Fig. 3(c), we observed a nominal cavity transmitted power of 1.8 mW. This value corresponds to an intracavity power of $P_{ic}(0) = 26$ W and a power at PD$_1$ of $P_{det}(0) = 110$ µW—recalling that OD = 1.2 for the NDF shown in Fig. 1(a). From the predicted standard error of the two-photon absorption coefficient [21, 23], we calculate for our observations—single ring-down events with noise-equivalent power $P_N \leq 120$ nW dominated by systematic sources—a minimum two-photon absorption coefficient of $1.6 \times 10^{-12}$ cm$^{-1}$ W$^{-1}$. This value is fourfold larger than the shot-noise limited sensitivity and tenfold larger than the technical-noise limited sensitivity (both calculated from Eq. (12) of Ref. [23], with our technical noise dominated by the transimpedance amplifier and oscilloscope card). Assuming an acquisition rate of $f_{acq} = \gamma_1/10 \approx 5.5$ kHz and eliminating the NDF, i.e., $P_{det}(0) = 1.8$ mW, we project [21] a shot-noise-equivalent two-photon absorption coefficient for our spectrometer of $\sigma(\gamma_2)/(c\sqrt{f_{acq}}) = 2.7 \times 10^{-15}$ cm$^{-1}$ W$^{-1}$ Hz$^{-1/2}$, were $\sigma(\gamma_2)$ is the standard error in $\gamma_2$. This result is equivalent to a 1 s N$_2$O detection limit of $\sigma_{N_2O}(1\text{ s}) = 7.9$ pmol mol$^{-1}$ Hz$^{-1/2}$.

A recent intercomparison of analytical instruments reported sensitivity limits for both $N_2O$ mole fraction and several of its relative isotopic abundances [36]. The study included laser-based gas analyzers operating near $\lambda$ = 4.53 μm which utilized linear absorption techniques like CRDS, off-axis integrated cavity output spectroscopy (OA-ICOS), and long pathlength quantum cascade laser absorption spectroscopy (QCLAS). For a gas sample with $N_2O$ mole fraction of 10 μmol mol$^{-1}$, comparable to our sample of 24.8 μmol mol$^{-1}$, they observed 1-s precision values within the range of $\sigma_{N_2O}(1\text{ s})$ = 1 nmol mol$^{-1}$ Hz$^{-1/2}$ to $\sigma_{N_2O}(1\text{ s})$ = 50 nmol mol$^{-1}$ Hz$^{-1/2}$. If achieved, our shot-noise limited projection for $\sigma_{N_2O}(1\text{ s})$ = 7.9 pmol mol$^{-1}$ Hz$^{-1/2}$ would out-perform the best linear absorption analyzer by 125-fold, thus illustrating the ultimate sensitivity of two-photon CRDS enabled by optical-electronic QCL stabilization. Further, two-photon CRDS promises intrinsic long-term stability by simultaneously measuring linear absorption and two-photon rates.

*Conclusions*—Cavity ring-down spectroscopy possesses several well-known advantages over conventional detection approaches. Specifically, it measures the photon decay rate which is easily related to the loss per unit length. Also, the passive decay signals are immune to laser intensity noise, and decay rates can be retrieved with shot-noise limited sensitivity [37] and metrology-grade accuracy [38]. We have demonstrated here that the high circulating power and counterpropagating laser fields characteristic of cavity ring-down spectroscopy can yield intense, velocity-class-indiscriminate, Doppler-free two-photon absorption spectra.

Improving upon our first measurements is relatively straightforward. Laser stabilization to an independent reference cavity will eliminate complications associated with disrupting our phase-locked loop to initiate decays. This will enable acquisition rates limited by the one-photon decay rate, and the inclusion of a high extinction ratio optical switch, e.g., an acousto-optic modulator, will reduce uncertainty in fitted two-photon decay rates. We also anticipate improved tunability over a larger dynamic range of intracavity powers. Further phase stabilization of the laser to an absolute frequency reference would improve spectrometer fidelity and enable longer integration times and an assessment of overall accuracy. Therefore, the extreme sensitivity limits predicted in Ref. [21] appear well within reach. Combined with the prospect of tunable stabilized lasers throughout the infrared spectral region—achieved by either the novel optical-electronic laser phase-locking approach demonstrated here or another more elaborate scheme, e.g., Refs. [18-20]—two-photon cavity ring-down spectroscopy as a tool at the vanguard of modern molecular physics is on the horizon.

*Acknowledgements*—We acknowledge technical assistance from Qingnan Liu (NIST) during optical resonator design, fabrication and assembly, and Christina E. Cecelski (NIST) and Jennifer Carney (NIST) for the loan of a certified gas sample of $N_2O$-in-air. Hélène M. Fleurbaey (NIST) commented on the manuscript. This work was funded by NIST, the National Research Council (NRC) Postdoctoral Research Associateships Program, and the University of Virginia. Certain commercial equipment is identified in this letter in order to specify the experimental procedure adequately. Such identification is not intended to imply recommendation or endorsement by NIST, nor is it intended to imply that the equipment identified is necessarily the best available for the purpose.